\documentclass[conference]{IEEEtran}
\usepackage{cite}
\usepackage{amsmath,amssymb,amsfonts}
\usepackage{algorithmic}
\usepackage{graphicx}
\usepackage{textcomp}
\usepackage{xcolor}
\usepackage{cuted}
\usepackage{stfloats}

\usepackage{lineno,hyperref}
\usepackage{color,xcolor,colortbl}
\usepackage{multirow}
\usepackage{array}
\usepackage[linesnumbered,ruled,vlined]{algorithm2e}
\SetKwInput{KwInput}{Input}
\SetKwInput{KwOutput}{Output}
\usepackage{bigstrut}
\usepackage{array}

\def\BibTeX{{\rm B\kern-.05em{\sc i\kern-.025em b}\kern-.08em
    T\kern-.1667em\lower.7ex\hbox{E}\kern-.125emX}}
\begin{document}

\title{On the Use of Power Amplifier Nonlinearity Quotient to Improve Radio Frequency Fingerprint Identification in Time-Varying Channels}

\author{\IEEEauthorblockN{Lu~Yang\IEEEauthorrefmark{1},
Seyit~Camtepe\IEEEauthorrefmark{2},
Yansong~Gao\IEEEauthorrefmark{2}, 
Vicky~Liu\IEEEauthorrefmark{3}, and
Dhammika~Jayalath\IEEEauthorrefmark{1}}
\IEEEauthorblockA{\IEEEauthorrefmark{1}Faculty of Engineering, Queensland University of Technology, Brisbane, Australia}
\IEEEauthorblockA{\IEEEauthorrefmark{2}Data61, CSIRO, Sydney, Australia}
\IEEEauthorblockA{\IEEEauthorrefmark{3}Faculty of Science, Queensland University of Technology, Brisbane, Australia}
\IEEEauthorblockA{Corresponding Author: Lu Yang (Email: l41.yang@hdr.qut.edu.au)}}

\maketitle

\begin{abstract}
Radio frequency fingerprint identification (RFFI) is a lightweight device authentication technique particularly desirable for power-constrained devices, e.g., the Internet of things (IoT) devices. Similar to biometric fingerprinting, RFFI exploits the intrinsic and unique hardware impairments resulting from manufacturing, such as power amplifier (PA) nonlinearity, to develop methods for device detection and classification. Due to the nature of wireless transmission, received signals are volatile when communication environments change. The resulting radio frequency fingerprints (RFFs) are distorted, leading to low device detection and classification accuracy. We propose a PA nonlinearity quotient and transfer learning classifier to design the environment-robust RFFI method. Firstly, we formalized and demonstrated that the PA nonlinearity quotient is independent of environmental changes. Secondly, we implemented transfer learning on a base classifier generated by data collected in an anechoic chamber, further improving device authentication and reducing disk and memory storage requirements. Extensive experiments, including indoor and outdoor settings, were carried out using LoRa devices. It is corroborated that the proposed PA nonlinearity quotient and transfer learning classifier significantly improved device detection and device classification accuracy. For example, the classification accuracy was improved by 33.3\% and 34.5\% under indoor and outdoor settings, respectively, compared to conventional deep learning and spectrogram-based classifiers.
\end{abstract}


\section{Introduction}
The rapid growth of the Internet of things (IoT) device population has sparked extensive demands on IoT security in recent years. Many security-critical IoT applications need more stringent security support~\cite{hassija2019survey}. Device authentication is one of the most important categories, which includes rogue device detection and the classification of registered devices~\cite{yang2017survey}. Traditionally, device authentication is achieved by public-key cryptography (PKC). However, the implemented public key algorithms are not optimal for IoT devices because they are computationally costly. Further, PKC generally requires a certification authority when sharing keys. The authority may not always be available, considering the large volume and wide-area deployment of IoT devices~\cite{xu2015device}.

A lightweight and reliable authentication technique is thus required for IoT security. Radio frequency fingerprint identification (RFFI) is a non-cryptographic authentication technique that attracted much research interest~\cite{riyaz2018deep,zhang2019physical,sankhe2020no,xie2021generalizable}. It exploits the intrinsic features brought by various hardware impairments resulting from imperfect manufacturing processes. The features manifested as slight distortions on transmitted signals. Like the biometric characteristics used for authentication, the subtle features are unique for different devices and hard to duplicate. Therefore, receivers can extract the features from received signals, followed by the verification with the pre-shared feature information for device authentication. The process does not involve computationally costly algorithms; hence, it consumes less energy and is suitable for power-constrained IoT devices.

An RFFI classifier is a machine learning model trained using radio frequency fingerprints (RFFs) for multi-class classification. Specifically, deep learning is leveraged as it minimizes the process of locating transient signal segments~\cite{merchant2018deep,das2018deep,peng2020deep,roy2020rfal,he2020cooperative}. It automatically extracts RFFs from received signals, making it the technique requiring minimal manual selection to train RFFI classifiers. For the network architecture, convolutional neural network (CNN) is mostly implemented for image recognition tasks, which makes it especially suitable for device fingerprinting~\cite{ding2018specific,sankhe2019oracle,rajendran2020injecting,jian2021radio,qian2021specific,li2022radionet}. Among the feature selection, in-phase and quadrature (IQ) samples~\cite{zhou2021robust}, FFT results~\cite{robyns2017physical,guo2021specific}, and spectrogram~\cite{shen2021radioJEAC} are widely studied. In~\cite{shen2021radioJEAC}, the spectrogram CNN model was shown to achieve the highest classification accuracy. Therefore, we adopt deep learning and spectrogram-based classifiers to benchmark proposed classifiers.

Due to the nature of wireless communications, RFFs are susceptible to environmental changes. Large-scale fading, multipath fading, and the Doppler effect affect wireless channels and modify received signal characteristics~\cite{shen2022towards,soltani2020rf,yang2022channel}. Traditional RFFs, e.g., spectrogram, extracted from the received signals are distorted and cannot be used for authentication~\cite{restuccia2019deepradioid,al2020exposing,al2021deeplora}. We propose using a power amplifier (PA) nonlinearity quotient to mitigate the wireless channel effects introduced by environmental changes. The PA nonlinearity quotient is generated by taking division on the frequency domain of two consecutive signals transmitted with different power. The division mitigates the wireless channel effects, and RFFI classifiers are trained to exploit the resulting RFFs.

Implementing environment-robust RFFs is limited when communication environments have many fast-moving objects because multipath fading and the Doppler effect mostly dominate wireless channels. Particularly, fast fading can happen when transmitters have low data rates, i.e., IoT devices. In~\cite{soltani2020more,al2021deeplora,shen2022towards}, data augmentation is implemented to alleviate the impact of fast fading by training classifiers under channels with simulated multipath fading and the Doppler effect. However, the simulations had no pre-knowledge of the real deployment environments and significantly increased the required disk and memory storage for training classifiers.

Transfer learning can be implemented to combine RFFs resulting from different wireless channels~\cite{sharaf2016authentication,wang2020radio,kuzdeba2021transfer}. Hence, distortions caused by multipath fading and the Doppler effect are acknowledged in device authentication. The required storage for transfer learning is less than data augmentation. Therefore, we implement transfer learning to alleviate the impact of fast fading. Specifically, a base classifier is trained with the original RFFs of the devices under test (DUTs); then, the classifier is retrained with the RFFs collected in real deployment environments.

This paper aims to design and validate an environment-robust RFFI system for IoT device authentication. The approach trains a classifier using the PA nonlinearity quotient. Transfer learning is adopted to alleviate the impact of fast fading and reduce training costs. Extensive experiments, including indoor and outdoor settings, were carried out using LoRa devices. The results show that the proposed PA nonlinearity quotient and transfer learning classifier significantly outperformed conventional deep learning and spectrogram-based classifiers. Our contributions are summarized as follows.

\begin{itemize}
    \item We formalized the PA nonlinearity quotient and demonstrated that it is independent of environmental changes. The improvements in rogue device detection and device classification are backed by experimental validation.
    \item We developed data collection of real deployment, including indoor and outdoor environments. Further, we implemented transfer learning using the data to alleviate the impact of fast fading. The approach reduced the disk and memory storage requirements for training. The resulting classifiers have pre-knowledge of the real deployment environments compared to the data augmentation approach.
    \item We designed an RFFI system that involves the PA nonlinearity quotient and transfer learning. Samples resulting from natural multipath fading and the Doppler effect were implemented to validate the system.
\end{itemize}


\section{Power Amplifier Nonlinearity Quotient}
The PA is an indispensable component in any wireless device, with the implementation to amplify low-power signals to high-power ones. It is inherently nonlinear~\cite{zhu2013challenges}. For low-power and narrowband systems, i.e., IoT devices, the PA is regarded as memoryless, meaning the nonlinear output depends only on the input at a particular time. The nonlinearity can be characterized by an amplitude/amplitude (AM/AM) function and an amplitude/phase (AM/PM) function. Several models have been proposed to formulate the functions~\cite{zhu2013challenges}.

Implementing PA nonlinearity for RFFI is widely studied in the literature~\cite{gong2020unsupervised,polak2011identifying,zhang2016specific,hanna2019deep,satija2019specific,li2022radio}. However, the implementation is often limited for static or semi-static channels. The RFFI performance drops significantly when communication environments change. We propose the PA nonlinearity quotient to design an environment-robust RFFI.

The signal of a narrowband system that reaches a receiver is given as
\begin{align*}
s(t)=h(\tau,t)\ast f\left [ x(t) \right ]+n(t),
\label{eq:signalModel}
\tag{1}
\end{align*}
where $x(t)$ is baseband signal, $h(\tau,t)$ is channel impulse response, $f[\cdot]$ denotes the nonlinear effect of hardware impairment at transmission power, and $n(t)$ is additive white Gaussian noise (AWGN). ``$\ast$'' denotes convolution operation.

When generating the PA nonlinearity quotient, two consecutive signals emitted with high and low transmission power correspondingly are received and developed an element-wise division on the frequency domain. The signal representation on the frequency domain is obtained through the short-time Fourier transform (STFT). The result of the STFT on the received signal is a matrix expressed as
\begin{flalign}
\boldsymbol{S}_p=
\begin{bmatrix}
S^{1,1}_p & S^{1,2}_p & \cdots & S^{1,M}_p \\
S^{2,1}_p & S^{2,2}_p & \cdots & S^{2,M}_p \\
\vdots & \vdots & \ddots & \vdots \\
S^{W,1}_p & S^{W,2}_p & \cdots & S^{W,M}_p
\end{bmatrix},
\label{eq:receivedspectrogram}
\tag{2}
\end{flalign}
where $p=\{h,l\}$ denotes high-power and low-power, respectively. The elements in the matrix are given as
\begin{align*}
S_{p}^{w,m}=\sum_{n=0}^{W-1}s_{p}\left [ n \right ]g\left [ n-mR \right ]e^{-j2\pi \frac{w}{W}n}\\
\text{for }w=1,2,...,W \text{ and }m=1,2,...,M,
\label{eq:STFT}
\tag{3}
\end{align*}
where $s_{p}\left [ n \right ]$ is the discrete signal received by the receiver with a sampling interval, $g\left [ n \right ]$ is the window function with length $W$, and $R$ is hop size. The experiments implement LoRa, hence $M$ is given by LoRa configurations as
\begin{align*}
M=\frac{K\cdot \frac{2^{SF}}{B}\cdot f_S-W}{R}+1,
\label{eq:Mvalue}
\tag{4}
\end{align*}
where $K$ is number of LoRa symbols, $SF$ is LoRa spreading factor, $B$ is bandwidth, and $f_S$ is sampling frequency. The configurations are discussed in Section~\ref{sec:ExperimentalSettings}. $W$ is 1024 and $R$ is 512. $M$ is calculated to be 319.

The STFT result of the high-power signal is expressed as~\eqref{eq:highpowerspectrogram}, where $X$ denotes the ideal spectrum of the transmitted signal, $H$ denotes the channel frequency response, and $F(\cdot)$ denotes the nonlinear hardware effect at the transmission power in the frequency domain. Only the preamble of the received signal is used to generate the PA nonlinearity quotient. The ideal spectrum of the low-power preamble is the same as the high-power one, i.e., $X^{w,m}=X^{w,M+m}$. Hence, the STFT result of the consecutive low-power signal is given as~\eqref{eq:lowpowerspectrogram}.

\begin{figure*}[b]
\begin{flalign}
\boldsymbol{S}_h=
\begin{bmatrix}
H^{1,1}F_{h}(X^{1,1}) & H^{1,2}F_{h}(X^{1,2}) & \cdots & H^{1,M}F_{h}(X^{1,M}) \\
H^{2,1}F_{h}(X^{2,1}) & H^{2,2}F_{h}(X^{2,2}) & \cdots & H^{2,M}F_{h}(X^{2,M}) \\
\vdots & \vdots & \ddots & \vdots \\
H^{W,1}F_{h}(X^{W,1}) & H^{W,2}F_{h}(X^{W,2}) & \cdots & H^{W,M}F_{h}(X^{W,M}) 
\end{bmatrix},
\label{eq:highpowerspectrogram}
\tag{5.1}
\end{flalign}
\end{figure*}

\begin{figure*}[b]
\begin{flalign}
\boldsymbol{S}_l=
\begin{bmatrix}
H^{1,M+1}F_{l}(X^{1,1}) & H^{1,M+2}F_{l}(X^{1,2}) & \cdots & H^{1,2M}F_{l}(X^{1,M}) \\
H^{2,M+1}F_{l}(X^{2,1}) & H^{2,M+2}F_{l}(X^{2,2}) & \cdots & H^{2,2M}F_{l}(X^{2,M}) \\
\vdots & \vdots & \ddots & \vdots \\
H^{W,M+1}F_{l}(X^{W,1}) & H^{W,M+2}F_{l}(X^{W,2}) & \cdots & H^{W,2M}F_{l}(X^{W,M}) 
\end{bmatrix}.
\label{eq:lowpowerspectrogram}
\tag{5.2}
\end{flalign}
\end{figure*}

By removing the significantly distorted preambles caused by fast-moving objects nearby and implementing transfer learning, we assume intense multipath fading and the Doppler effect are mitigated. Slow fading mostly dominates the wireless channels. Therefore, the channel frequency response does not change significantly during one packet duration, i.e., $H^{w,m}\approx H^{w,M+m}$. The result of the element-wise division of received signals on the frequency domain ($\boldsymbol{Q}$) is given as
\begin{flalign}
\boldsymbol{Q}=\boldsymbol{S}_h ./ \boldsymbol{S}_l=
\begin{bmatrix}
\frac{F_{h}(\boldsymbol{X^{1}})}{F_{l}(\boldsymbol{X^{1}})} & \frac{F_{h}(\boldsymbol{X^{2}})}{F_{l}(\boldsymbol{X^{2}})} & \boldsymbol\cdots & \frac{F_{h}(\boldsymbol{X^{M}})}{F_{l}(\boldsymbol{X^{M}})} \\
\end{bmatrix},
\label{eq:PAnonlinearityquotient}
\tag{6}
\end{flalign}
where ``$./$'' denotes the element-wise division operation and $\boldsymbol{X^m}=[X^{1,m} \quad X^{2,m} \quad \cdots \quad X^{W,m}]^T$. No channel frequency response ($H$) is present in $\boldsymbol{Q}$. The proposed environment-robust RFFI can be developed exploiting the PA nonlinearity quotient, which is $\boldsymbol{Q}$ in dB scale, expressed as
\begin{align*}
\widetilde{\boldsymbol{Q}}=10\log_{10}(|\boldsymbol{Q}|^2).
\label{eq:QdB}
\tag{7}
\end{align*}

\section{Experiments}

\subsection{Experimental Settings}\label{sec:ExperimentalSettings}
\begin{figure}[t]
\centering
\includegraphics[width=0.42\textwidth]{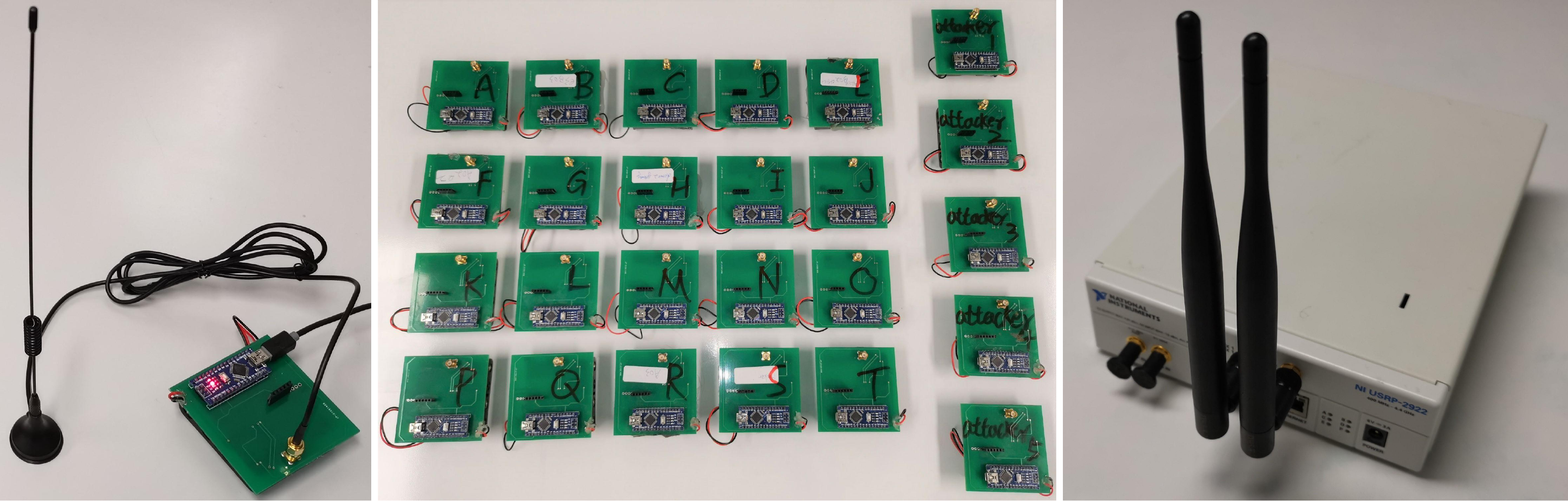}
\caption{Devices used in the experiments. Left: a DUT in operation. Middle: 20 DUTs as legitimate devices and 5 DUTs as rogue devices. Right: a USRP-2922 platform as the receiver.}
\label{fig:devices}
\end{figure}

The experiments implemented 25 Arduino Nano-controlled LoRa SX1276 modules with the same circuit design and specifications as DUTs. 20 DUTs were randomly selected as legitimate devices (DUT: ``A" to ``T"), and 5 DUTs were selected as rogue devices (DUT: ``Attacker 1" to ``Attacker 5"). The device configurations are given in Table~\ref{tab:config}. The LoRaWAN protocol supports 125~kHz, 250~kHz, and 500~kHz bandwidths, while LoRa supports bandwidths ranging from 7.8~kHz to 500~kHz. The proposed RSSI system does not focus on specific protocols. Therefore, a bandwidth of 62.5~kHz was used to reduce packet loss and maintain high throughputs. A universal software radio peripheral (USRP) platform with a 1~MS/s sampling frequency ($f_S$) was used to collect RF samples. Fig.~\ref{fig:devices} shows the devices used in the experiments.

\begin{table}[t]
  \footnotesize
  \centering
  \caption{DUT Configurations}
    \begin{tabular}{|c|c|c|c|c|}
    \hline
    \begin{tabular}[c]{@{}c@{}}Carrier\\Frequency\end{tabular} & \begin{tabular}[c]{@{}c@{}}Bandwidth\\($B$)\end{tabular} & \begin{tabular}[c]{@{}c@{}}Transmission\\Power (h/l)\end{tabular} & \begin{tabular}[c]{@{}c@{}}Spreading\\Factor ($SF$)\end{tabular} & \begin{tabular}[c]{@{}c@{}}Coding\\Rate\end{tabular} \bigstrut\\
    \hline
    915~MHz & 62.5~kHz & 17/10~dBm & 10 & 4/5 \bigstrut\\
    \hline
    \end{tabular}
  \label{tab:config}
\end{table}

\begin{figure}[t]
\centering
\includegraphics[width=0.42\textwidth]{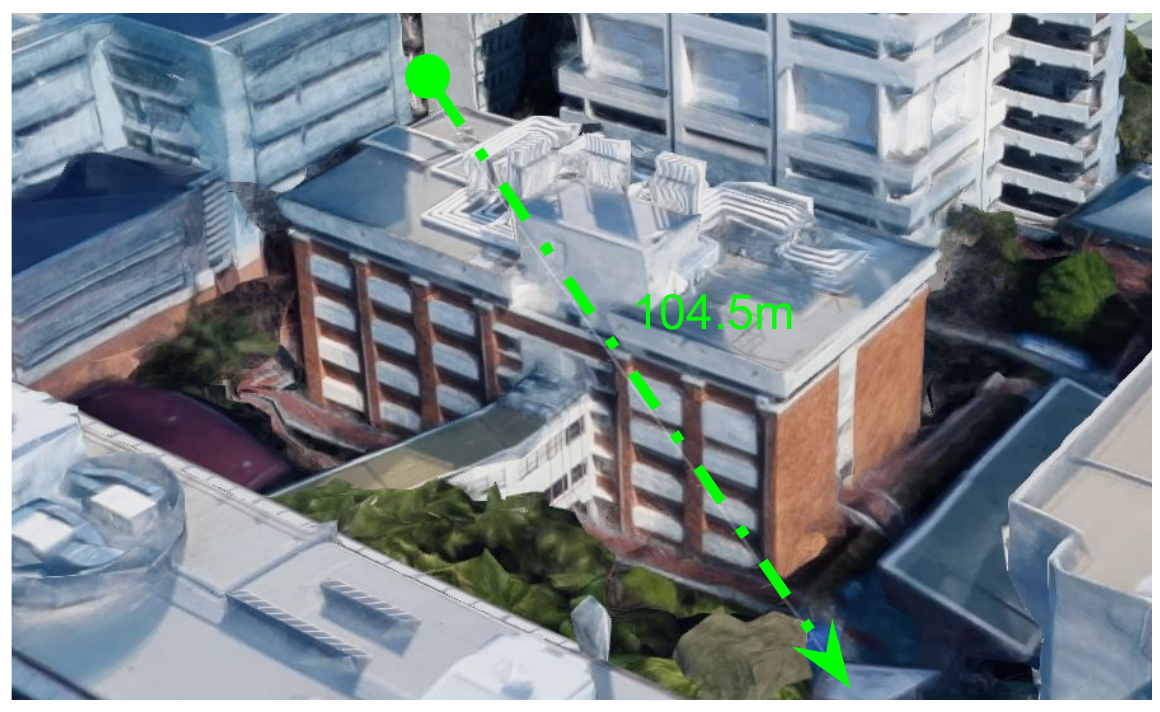}
\caption{Outdoor experimental environment.}
\label{fig:outdoorEn}
\end{figure}

The data collection was developed in three environments.
\begin{itemize}
    \item \textbf{Anechoic chamber:} the collection of channel effect-free RFFs for training the base classifier required by transfer learning was carried out in the anechoic chamber on the top floor of the QUT GP campus S-block building. DUTs were placed 3 meters away from the USRP platform. The anechoic chamber was designed to absorb multipath signals. Therefore, RF samples collected in the environment can generate the PA nonlinearity quotient without the impact of multipath fading and the Doppler effect.
    \item \textbf{Indoor:} DUTs were placed in an office room for the indoor setting. The USRP platform was placed in the adjacent room, and DUT signals traveled through a wall. People were freely walking in the office during the data collection. The environment was considered to have moderate multipath fading and a slight Doppler effect.
    \item \textbf{Outdoor:} in the outdoor setting, DUTs were placed 104.5 meters away from the USRP platform, as shown in Fig.~\ref{fig:outdoorEn}. Buildings blocked the line of sight, and people freely walked in the environment. The outdoor environment was considered to have more significant multipath fading and the Doppler effect than the indoor environment.
\end{itemize}

The DUTs transmit packets with alternating high-power and low-power modes, and the USRP platform passively receives the packets in the data collection. More than 2800 packets were collected for each DUT within one hour. Hence, more than 8400 packets were collected for each DUT in all three experimental environments.

\subsection{Data Preprocessing}
\begin{figure}[t]
\centering
\includegraphics[width=0.45\textwidth]{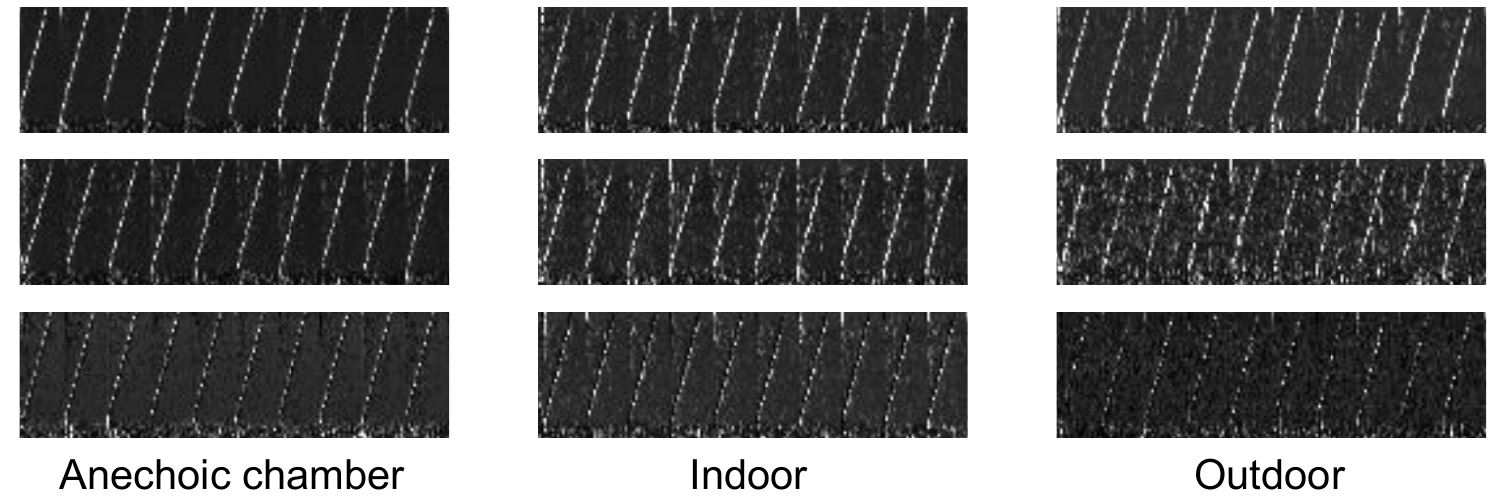}
\caption{Spectrogram and PA nonlinearity quotient of a DUT in the experiments. Top: high-power preamble spectrogram. Middle: low-power preamble spectrogram. Bottom: PA nonlinearity quotient, $\widetilde{\boldsymbol{Q}}$ in~\eqref{eq:QdB}.}
\label{fig:fingerprintIMG}
\end{figure}

The data preprocessing includes synchronization, preamble extraction, normalization, and the PA nonlinearity quotient generation. The packets collected indoors and outdoors are required to go through the distorted preamble removal process before generating the PA nonlinearity quotient.
\begin{enumerate}
    \item Synchronization: transmission power does not impact the data rate. Hence, the time-on-air for the DUT packets stays unchanged for the high-power and low-power transmission. The starting points of the packets are marked and used for synchronization to avoid inaccurate preamble extraction.
    \item Preamble extraction: preambles are payload-independent and have no software-defined features such as MAC addresses. Therefore, the intrinsic hardware features in the preamble symbols are the desirable source for RFFI. The preamble length is a flexible configuration for LoRa, with a minimum value of ten symbols. To study the worst-case scenario, we set and extracted ten symbols for one preamble per packet in the experiments.
    \item Normalization: the process normalizes the received signal magnitude to remove the device-specific DC offset by dividing the root mean square. The PA nonlinearity feature is unaffected.
    \item Distortion removal and PA nonlinearity quotient generation: we introduce Algorithm~\ref{algorithm:generation} to remove the severely distorted preambles caused by fast-moving objects nearby. The correlation between the high-power and low-power spectrogram should stay the same since PA nonlinearity is only affected by the input power~\cite{zhang2021radio}. The distorted preambles can be found by comparing the correlation of the channel-affected spectrogram to the correlation of the anechoic chamber spectrogram. The distortion is considered severe and can be removed if the difference is over a tolerance ($\theta=0.2$ implemented in experiments). After the distortion removal, an element-wise division on the frequency domain is developed to generate the PA nonlinearity quotient. Fig.~\ref{fig:fingerprintIMG} shows the collected preamble spectrogram and the PA nonlinearity quotient generated by a DUT.
\end{enumerate}

\begin{algorithm}[t]
\footnotesize
\DontPrintSemicolon
  \KwInput{$\boldsymbol{S}_{h,k},\boldsymbol{S}_{l,k}$ \quad $\%$STFT results of indoor or outdoor preambles ($k=indoor \: or \: outdoor$)}
  \KwInput{$\boldsymbol{S}_{h,c},\boldsymbol{S}_{l,c}$ \quad $\%$STFT results of anechoic chamber preambles}
  \KwInput{$\theta$ \quad $\%$ Tolerance}
  \KwOutput{$\widetilde{\boldsymbol{Q}}$ \quad $\%$ PA nonlinearity quotient}
  $\rho_k=corr\{max(\boldsymbol{S}_{h,k}),max(\boldsymbol{S}_{l,k})\}$\\
  $\rho_c=corr\{max(\boldsymbol{S}_{h,c}),max(\boldsymbol{S}_{l,c})\}$\\
  $\rho_d=\|\rho_c-\rho_k\|$\\
  \eIf{$\rho_d\leq\theta$}
  {$\left. \boldsymbol{Q}=\boldsymbol{S}_{h,k} .\middle/ \boldsymbol{S}_{l,k} \right.$  \quad $\%$ Element-wise division \\
  $\widetilde{\boldsymbol{Q}}=10\log_{10}(|\boldsymbol{Q}|^2)$}
  {$remove \: \boldsymbol{S}_{h,k},\boldsymbol{S}_{l,k}$}
\caption{Distortion Removal and PA Nonlinearity Quotient Generation.}
\label{algorithm:generation}
\end{algorithm}

\subsection{Analytical Metrics}
Device authentication exploiting RFFI involves two essential parts: device classification and rogue device detection. The classification accuracy and receiver operating characteristic (ROC) curve are implemented to evaluate the device classification and rogue device detection performance, respectively.
\subsubsection{Classification Accuracy}
The classification accuracy is defined as the number of correctly classified RFFs divided by the total number of tested RFFs. The results are obtained from the confusion matrix after developing classification tests.
\subsubsection{ROC Curve}
The rogue device detection was studied as binary classification in the experiments. The output values of the softmax function are compared to a threshold. The RFFs associated with the output values smaller than the threshold will be considered unauthorized. Since the threshold is configurable, it is hard to use a detection rate to analyze classifiers' performance. We adopted the ROC curve in the binary classifier study to overcome this. For each class of a classifier, ROC analysis applies threshold values in [0,1] to calculate the true-positive rate (TPR) and the false-positive rate (FPR) for the outputs generated by each threshold. The area under the ROC curve (AUC) is the integral of a ROC curve with respect to FPR. The value of AUC is in the range of 0 to 1. A larger AUC indicates better classifier performance. In our experiments, a larger AUC indicates that the classifier is more capable of detecting rogue devices. A micro-averaging method is applied to generate the averaged AUC and ROC curves to analyze the rogue device detection for all classes.

\section{Classifier Architecture}
The architecture of the PA nonlinearity quotient and transfer learning classifier is summarized in Table~\ref{table:ClassifierArchitecture}. It consists of three convolution layers with 8, 16, and 32 $3\times3$ filters, respectively. A batch normalization layer and the rectified linear unit (ReLU) activation follow each convolution layer. After the activation, a $2\times2$ max pooling layer with stride 2 is implemented. The output of the last ReLU activation is fed to a fully connected layer. An output layer with softmax function is implemented last to produce vectors of probabilities of outputs. The PA nonlinearity quotient is resized to $256\times256$ with 8-bit depth to go to the input layer. Adam optimizer is implemented to reduce the losses. The mini-batch size is 32. The initial training rate is 0.005 and remains unchanged.

Transfer learning retrains a pre-trained classifier on new datasets. In the experiments, the convolution layers of the pre-trained classifier recognize generic RFF patterns. We replaced the fully connected and output layers with new layers. For fine-tuning the transferred classifier, the training rate was configured to 0.0001, and the learning rate factor of the new layers was configured to 20.

\begin{table}[t]
\footnotesize
\centering
\caption{Layers, Parameters, and Activation of the Proposed Classifier}
\label{table:ClassifierArchitecture}
\begin{tabular}{|c|c|c|c|}
\hline
Layer           & Dimension             & Parameters    & Activation\\\hline
Input           & $256\times256$        & ---           & ---\\       \hline
Convolution, BN & $8\times(3\times3)$   & 80, 16        & ReLU\\      \hline
MaxPooling      & $2\times2$            & ---           & ---\\       \hline
Convolution, BN & $16\times(3\times3)$  & 1168, 32      & ReLU\\      \hline
MaxPooling      & $2\times2$            & ---           & ---\\       \hline
Convolution, BN & $32\times(3\times3)$  & 4640, 64      & ReLU\\      \hline
FullyConnected  & 20                    & 2304020       & SoftMax\\   \hline
\end{tabular}
\end{table}

\section{Results and Discussion}

\subsection{Device Classification}

\begin{figure}[t]
\centering
\includegraphics[width=0.44\textwidth]{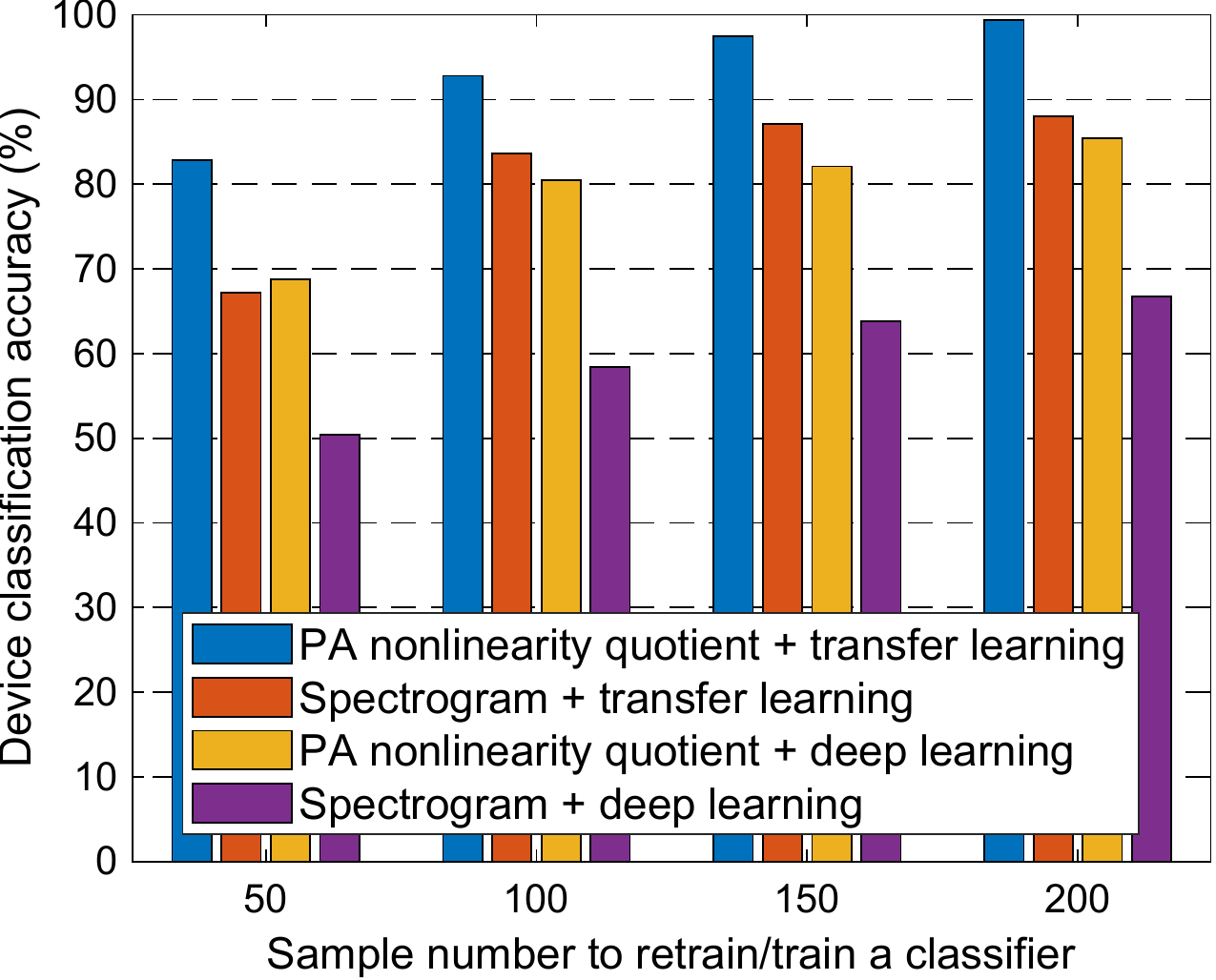}
\caption{Device classification results of the indoor experiments.}
\label{fig:indoorClassification}
\bigbreak
\includegraphics[width=0.44\textwidth]{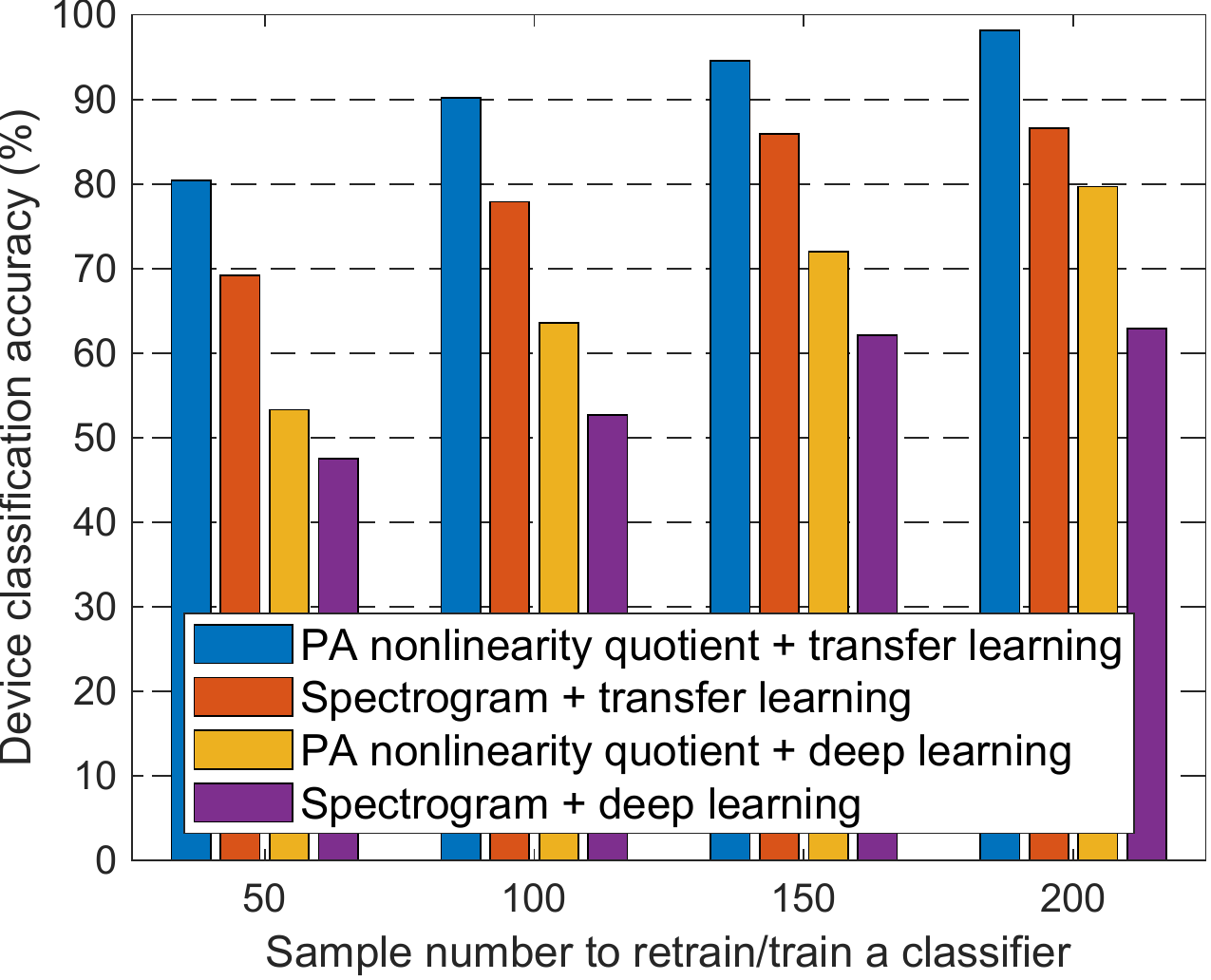}
\caption{Device classification results of the outdoor experiments.}
\label{fig:outdoorClassification}
\end{figure}

The base classifier was trained firstly using complete legitimate device (DUT: ``A" to ``T") datasets in the anechoic chamber. Smaller training sets, including 50, 100, 150, and 200 packets, were randomly selected for each legitimate device from the indoor and outdoor datasets to implement the transfer learning. The conventional deep learning and spectrogram-based classifiers were trained as the comparison. The same test sets, including more than 1000 packets per DUT, were implemented to validate the proposed PA nonlinearity quotient and transfer learning classifier and the deep learning and spectrogram-based classifier. No training set packets were used in the test sets.

Fig.~\ref{fig:indoorClassification} shows the device classification results of indoor experiments. The proposed PA nonlinearity quotient and transfer learning classifier outperformed the conventional deep learning and spectrogram-based classifier with an improvement of $33.3\%$ average classification accuracy. More training packets lead to higher classification accuracy. The highest accuracy is $99.4\%$, with 200 packets retraining the base classifier. The PA nonlinearity quotient improved the average classification accuracy by $19.4\%$ compared to the spectrogram-based classifier.

Fig.~\ref{fig:outdoorClassification} shows the classification results of outdoor experiments. The proposed PA nonlinearity quotient and transfer learning classifier outperformed the conventional deep learning and spectrogram-based classifier with an improvement of $34.5\%$ average classification accuracy. The PA nonlinearity quotient improved the average classification accuracy by $10.9\%$ compared to the spectrogram-based classifier.

Table~\ref{table:DeviceClassificationComparison} compares device classification performance among the proposed classifier and recent notable works in literature. The PA nonlinearity quotient and transfer learning classifier achieved high device classification accuracy while requiring fewer training samples and reducing the disk and memory storage requirements.
\begin{table}[t]
\footnotesize
\centering
\caption{Device Classification Comparison With Notable Works}
\label{table:DeviceClassificationComparison}
\begin{tabular}{|c|c|c|c|c|}
\hline
Work & \begin{tabular}[c]{@{}c@{}}Experimental\\Environment\end{tabular} & \begin{tabular}[c]{@{}c@{}}No. of\\Devices\end{tabular} &\begin{tabular}[c]{@{}c@{}}Training Samples\\(Per Device)\end{tabular} & \begin{tabular}[c]{@{}c@{}}Classification\\Accuracy\end{tabular} \\\hline
Ours                  & \begin{tabular}[c]{@{}c@{}}Indoor\\Outdoor\end{tabular} & 20 & 200 & \begin{tabular}[c]{@{}c@{}}99.4\%\\98.2\%\end{tabular} \\\hline
\cite{shen2022towards}   & Indoor   & 30    & 100   & 98.4\%    \\\hline
\cite{yu2019robust}      & Indoor   & 54    & 698   & 84.6\%    \\\hline
\cite{xing2022design}    & Indoor   & 7     & 800   & 99.0\%    \\\hline
\end{tabular}
\end{table}
\subsection{Rogue Device Detection}
The training sets to retrain the base classifier included 100 randomly selected packets per DUT for studying the rogue device detection for the proposed classifier. Deep learning and spectrogram-based classifiers were trained for comparison. The test sets included more than 1000 packets per DUT and more than 1000 packets for each rogue device (DUT: "Attacker~1" to "Attacker~5"). No training set packets were used in the test sets.

\begin{figure}[t]
\centering
\includegraphics[width=0.4\textwidth]{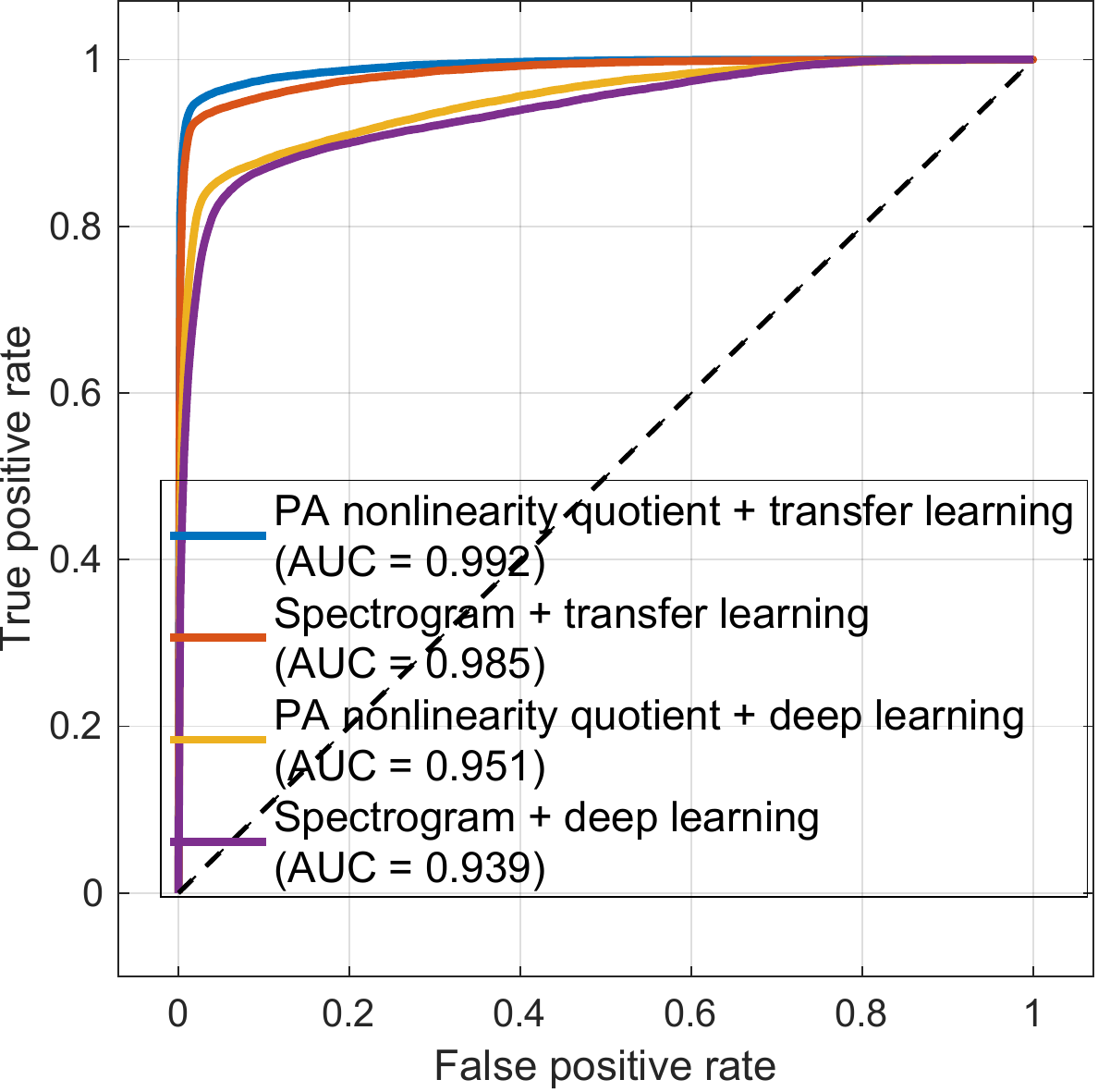}
\caption{ROC curves of rogue device detection in the indoor experiments.}
\label{fig:indoorROC}
\bigbreak
\includegraphics[width=0.4\textwidth]{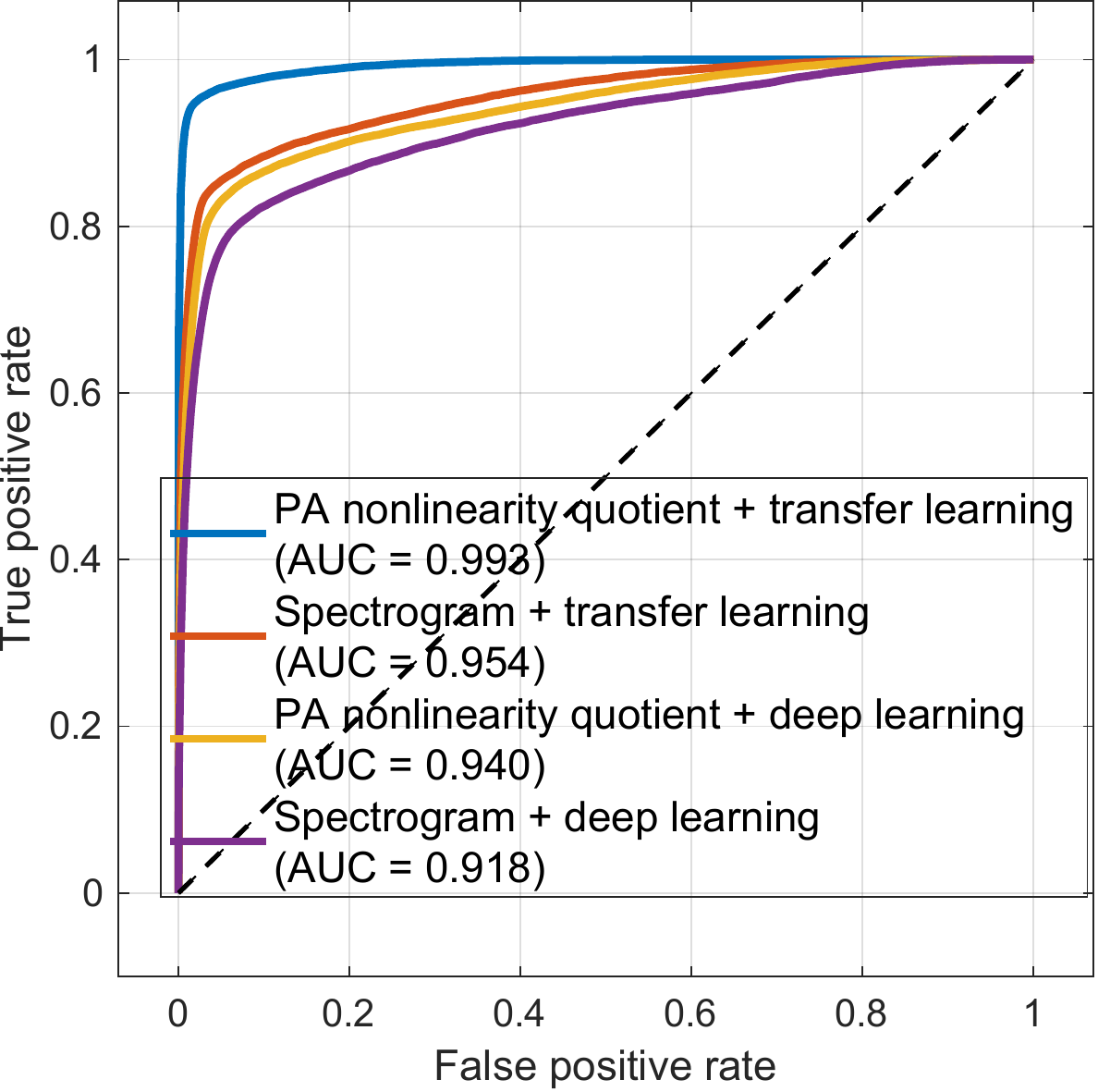}
\caption{ROC curves of rogue device detection in the outdoor experiments.}
\label{fig:outdoorROC}
\end{figure}

Fig.~\ref{fig:indoorROC} shows the ROC curves for the indoor experiments. The proposed PA nonlinearity quotient and transfer learning classifier outperformed the deep learning and spectrogram-based classifier, with an AUC value of 0.992 compared to 0.939. Fig.~\ref{fig:outdoorROC} shows the outdoor experiment results. Similar to the indoor experiments, the proposed classifier improved the AUC significantly. The PA nonlinearity quotient was more robust to environmental changes than the spectrogram, with larger AUC values in the indoor and outdoor experiments.

\section{Conclusion}
In this paper, we investigated the technique to make RFFI resilient to environmental changes. We proposed the PA nonlinearity quotient and transfer learning classifier that mitigates channel effects to enhance the RFFI implementation for device classification and rogue device detection. Extensive experiments, including indoor and outdoor settings, were developed to evaluate the proposed classifier. The experiment results demonstrated that the proposed classifier significantly improved classification accuracy and rogue device detection for RFFI. The PA nonlinearity quotient outperformed the spectrogram to enhance RFFI in indoor and outdoor settings.

\bibliographystyle{IEEEtran}
\bibliography{bib.bib}

\end{document}